\documentclass[fleqn,oneside]{article}    
\usepackage{espcrc2}
\title{On helicity and spin on the light cone}
\author{A. Krassnigg\address{Institut f\"ur Theoretische Physik, 
        Karl-Franzens-Universit\"at, A--8010 Graz, Austria}
        and
        H.C. Pauli\address{Max--Planck Institut f\"ur Kernphysik, 
        D--69029 Heidelberg, Germany,} }
\begin{document}
\begin{abstract}
Starting from a one--body front--form equation 
with Lepage-Brodsky spinors we show, with a fair amount of
new technology, how an integral equation in standard momentum space 
with Bj\o rken-Drell spinors can be obtained.
The integral equation decouples for singlets and triplets.
\hfill                       HCP              2Oct/20 Nov 2001 
\vspace{1pc}
\end{abstract}
\maketitle
\section{Introduction}
We address to Eq.(96) of \cite{Pau98},
\begin{eqnarray} 
    M^2\psi_{\lambda_1\lambda_2}(x,\vec k_{\!\perp}) 
    = \sum _{ \lambda_1',\lambda_2'} 
    \!\!\int\!\! dx' d^2 \vec k_{\!\perp}' 
\nonumber\\ \times   
    \ U_{\lambda_1\lambda_2;\lambda_1'\lambda_2'}
    (x,\vec k_{\!\perp};x',\vec k_{\!\perp}')
    \ \psi_{\lambda_1'\lambda_2'}(x',\vec k_{\!\perp}')   
\nonumber\\ +
    \left[ 
    \frac{m^2_{1} + \vec k_{\!\perp}^{\,2}}{x} +
    \frac{m^2_{2} + \vec k_{\!\perp}^{\,2}}{1-x}  
    \right]
    \psi_{\lambda_1\lambda_2}(x,\vec k_{\!\perp})  
,\label{eq:1}\end{eqnarray} 
a one-body integral equation with the kernel
\begin{eqnarray} \nonumber
    U_{\lambda_1\lambda_2;\lambda_1'\lambda_2'}
    (x,\vec k_{\!\perp};x',\vec k_{\!\perp}') = -
    \frac{4m_1m_2}{3\pi^2} 
\\ \times
    \frac{\overline\alpha(Q)}{Q^2}R(Q)
    \frac{S_{\lambda_1\lambda_2;
    \lambda_1'\lambda_2'}(x,\vec k_{\!\perp};x',\vec k_{\!\perp}')}
    {\sqrt{ x(1-x) x'(1-x')}}
,\end{eqnarray} 
and refer to \cite{Pau01d} and \cite{Pau01e} 
for more background information.
Here, $M ^2$ is the wanted eigenvalue of the 
invariant mass squared operator, 
with associated eigenfunction $\psi\equiv\Psi_{q\bar q}$. 
It is the probability amplitude 
for finding in the $q\bar q$--space 
a quark with the effective (constituent) mass $m_1$, 
longitudinal momentum fraction $x$, 
transversal momentum $\vec k_{\!\perp}$ 
and helicity $\lambda_{1}$,
and correspondingly for the anti--quark with $m_2$, 
$1-x$, $-\vec k_{\!\perp}$ and $\lambda_{2}$.

The spinor factor $S_{\lambda_1\lambda_2;\lambda_1'\lambda_2'}$ 
is tabulated explicitly in \cite{Pau00d}.
It is defined in terms of Lepage-Brodsky spinors \cite{BPP98,Pau00d}, 
\begin{eqnarray}
\begin{array} {@{}l@{}l@{}l@{}}
   &S_{\lambda_1\lambda_2;\lambda_1'\lambda_2'}
   (x,\vec k_{\!\perp};x',\vec k_{\!\perp}')=
\\
   &\left[\overline u(k_1,\lambda_1)\gamma^\mu
   u(k_1',\lambda_1')\right] 
   \left[\overline u (k_2,\lambda_2)\gamma_\mu
   u(k_2',\lambda_2')\right].  
\end{array}
\end{eqnarray}
But contrary to \cite{Pau00d}, the $u$-spinors are normalized here:
\(
   \overline u (k,\lambda) u (k,\lambda') = 
   \delta_{\lambda\lambda'}
.\)
Due to the helicity indices, the
one--body equation in Eq.(\ref{eq:1})
is a set of four coupled integral equations
in the three momentum components $x$ and $\vec k_{\!\perp}$. 

Our aim is to convert Eq.(1) into a set of integral equations
in usual momentum space.

\section{Helicity and spin}
The Lepage-Brodsky spinors $u(p,\lambda)$,
as tabulated in the appendix of Ref.~\cite{BPP98}, are:
\begin{equation} 
   u^{\scriptstyle LB}(p,\lambda) \! = \! 
   \frac{1}{\sqrt{4mp^+}}
   \left(\begin{array}{cc}
   p^+ \!+\!m & -p_l \\ p_r & p^+ \!+\!m \\ p^+ \!-\!m & p_l \\ 
   p_r & -p^+ \!+\!m  
   \end{array}\right) 
\end{equation}
for $\lambda:\;\uparrow\downarrow$,
with $p_r \equiv p_x+ip_y$ and $p_l \equiv p_x-ip_y$.
They differ from the Bj\o rken--Drell spinors \cite{BjD64},
\begin{equation} 
   u^{\scriptstyle BD}(p,s) \! \! = \! 
   \frac{1}{\sqrt{2m(E\!+\!m)}} \! \! 
   \left(\begin{array}{cc}
   E\!+\!m & 0 \\ 0 & E\!+\!m \\ p_z & p_l \\ p_r & -p_z 
   \end{array}\right) \! \! 
\label{eq:5}\end{equation} 
for $s:\;\uparrow\downarrow$, with 
$E\equiv E(p) = \sqrt{m^2+\vec p_{\!\perp}^{\;2}+p_z^2}$.

Both spinors are solutions to the same equation, 
the free Dirac equation
\(
(/\!\!\!p-m)\, u(p,\lambda)=0
\)
and are normalized in the same way. 
Hence, they must be linear superpositions of each other:
\begin{equation}
   u ^{BD} _\alpha (p,s)= \sum _\lambda u ^{LB} _\alpha (p,\lambda)
   \ \omega_{\lambda s } 
.\end{equation}
The transformation matrix can be constructed easily by projection with
the appropriate adjoint spinor. We get
\begin{equation}
   \omega_{\lambda s } = \sum_{\alpha}
   \overline u^{LB}_\alpha(p,\lambda)\ u^{{BD}}_\alpha(p,s) .
\end{equation}
Calculating the four overlap matrix elements explicitly
with the corresponding spinors yields:
\begin{equation}
  \omega_{\lambda s } \! = \! \! 
  \left(\begin{array}{@{}cc@{}} m+p^+ & p_l \\ -p_r & m+p^+ \end{array}\right) 
  \frac{1}{\sqrt{2p^+(E+m)}} ,  
\end{equation}
with the rows labeled by the helicity $\lambda =\uparrow\downarrow$ 
and the columns by the spin $s=\uparrow\downarrow$. 
This transformation is unitary. It is called a Melosh transformation 
\cite{Mel74,Dzi87,AhS92}.
The matrix $\omega_{\lambda s}$ has dimension 
$2\times 2$ in our case (fermions). 
For higher spins these matrices are larger; they have dimension
$(2j+1)\times(2j+1)$.
The unitarity relation
\(
  \omega \omega^\dagger = \omega^\dagger\omega = 1
\)
can be verified easily by
\begin{eqnarray}
  \hspace{-1em}\sum _{s}
  \omega_{\lambda s} \omega_{s \lambda'}^\star = 
  \delta _{\lambda \lambda'} =
  \frac{1}{2p^+(E+m)} 
\times\phantom{,}\nonumber\\ 
  \left(\begin{array}{@{}c@{}c@{}}  m+p^+ &  p_l \\ -p_r & m+p^+ 
  \end{array}\right)
  \left(\begin{array}{@{}c@{}c@{}}  m+p^+ & -p_l \\ p_r & m+p^+ 
  \end{array}\right)
,\end{eqnarray}
since $(p^+ +m)^2+ p_l p_r = 2p^+(E+m)$.

The spinors appear in bilinear combinations.
It is therefore convenient to define the $4\times 4$ matrix
\begin{equation} 
   \Omega _{\lambda_1\lambda_2;s _1 s_2} = 
   \omega _{\lambda_1 s _1}  
   \omega _{\lambda_2 s _2}   
\end{equation}
as the unitary direct product of the $\omega$'s. 
In consequence one can apply a unitary transformation 
to generate a new spinor factor 
\begin{eqnarray} 
   \widetilde  S _{s_1s_2;s_1's_2'} =
   \left(\Omega ^\dagger S \Omega \right) _{s_1s_2;s_1's_2'} = 
\nonumber\\  
   \left[\overline u(k_1,s_1)\gamma^\mu
   u(k_1',s_1')\right] 
   \left[\overline u (k_2,s_2)\gamma_\mu
   u(k_2',s_2')\right] 
,\label{eq:11}\end{eqnarray}
which looks like $S$ except that 
Lepage--Brodsky are replaced by Bj\o rken--Drell spinors.

The BD--spinors of Eq.(\ref{eq:5}) transform under spatial rotations 
like a half-integer representation of the rotation group 
$\vert t\ s\rangle$ with spin $t=\frac{1}{2}$ 
and projection $s=\pm\frac{1}{2}$ \cite{Edm64}.
Two such spin states can be coupled to a state 
with total spin $T=0$ or $T=1$.
Particularly, the function
\[
   W_{T S} =  
   \sum_{s_1,s_2} \langle t_1s_1t_2s_2 \vert T S\rangle
   u (k_1,s_1) u (k_2,s_2)
,\]
with $t_1=t_2=\frac{1}{2}$,
transforms like a function of $k_1$ and $k_2$ 
with total spin $T$ and projection $S=-T,-T+1,\dots,T$. 
The Clebsch-Gordan coefficients 
$\langle t_1s_1t_2s_2 \vert T S\rangle\equiv C_{s_1 s_2 ; T S }$
are real in the Condon-Shortley phase convention \cite{Edm64} 
and define an unitary transformation.
We use it to define yet a third spinor factor $\overline S$
\begin{eqnarray}
 \begin{array}{llllll} 
   &\overline S _{T S;T' S'} &=& 
   \big( C ^\dagger \widetilde S C\big) _{T S;T' S'}  
\\ & 
   &=&
   \big( C ^\dagger \Omega ^\dagger S \Omega C\big) _{T S;T' S'}    
 \end{array}
\end{eqnarray}
and proceed.

\section{Transforming the integral equation}
In comparing Lepage--Brodsky and Bj\o rken--Drell
spinors we have used tacidly the relation between the 
instant form $p_z$ and the front form 
$p ^+ = p_z + \sqrt{m^2+\vec p_{\!\perp}^{\;2}+p_z^2}$.
This is correct, since the Dirac spinors are defined
only for free particles.
Dealing with two particles, and two spinors $u (k_1,s_1)$ and 
$u (k_2,s_2)$, one can assume that they have opposite three-momenta
$\vec k _1= -\vec k _2 = \vec k$.
Thus, their $k ^+$ and total 
$K ^+ = k _1^+ + k _2^+$ are related to $k_z$ 
(with $\vec k= (\vec k_{\!\perp},k_z)$).
One can introduce even a momentum fraction $x=k _1^+/K ^+$
and relate it to $k_z$ by
\begin{equation}
    x(k_z)  = \frac{E_1+k_z}{E_1+E_2}
\label{eq:14}\end{equation}
with
\(
    E_{1,2} \equiv E_{1,2} (k) =
    \sqrt{m^{\,2}_{1,2}+ k_z^2 + \vec k_{\!\perp}^{\,2}} 
\).
But here is a subtle point: The so defined $x$ is conceptually
different from the longitudinal momentum fraction in Eq.(\ref{eq:1})
which is $x=k_1^+/P^+$. 
The $P^+$ contains the interaction \cite{BPP98},
the $K^+$ does not.

However, there is one common aspect: While $k_z$ varies from 
$-\infty$ to $+\infty$, the $x(k_z)$ in Eq.(\ref{eq:14}) 
varies from $0$ to $1$.
It has thus the same domain of validity as the momentum fractions
of Eq.(\ref{eq:1}). Eq.(\ref{eq:14}) can be used therefore to transform
integration variables in Eq.(\ref{eq:1})
from $(x,\vec k_{\!\perp})$ to $\vec k$, 
such that all three components of $\vec k$ have the same domain. 
In numerical work this is more than convenient. 
The transformation in Eq.(\ref{eq:14}) has thus been used tacidly 
in all past numerical work. 
New is here, that we use Eq.(\ref{eq:14}) explicitly
to transform Eq.(\ref{eq:1}).

Transforming variables, the Jacobian is \cite{Pau00d}
\begin{equation}
    dx = \frac {x(1-x)} {A (k)}\ \frac {dk_z}{m_r} 
.\end{equation}
The reduced mass $m_r$ takes care of the dimensions since
$A (k)$ is dimensionless, 
\begin{equation}
    \frac{1}{A (k)} = \frac{m_r }{E_r} = m_r 
    \left(\frac{1}{E_1} + \frac{1}{E_2}\right)
.\end{equation}
The diagonal part of Eq.(\ref{eq:1}) becomes
\begin{equation}
    \frac{ m^2_{1} + \vec k_{\!\perp}^{\,2}}{x} +
    \frac{ m^2_{2} + \vec k_{\!\perp}^{\,2}}{1-x}  
    =  \left(E _1(k) + E _2(k)\right)^2 
.\end{equation}
Substituting all of that in Eq.(\ref{eq:1}) 
gives at first 
\begin{eqnarray} 
    M^2 \psi_{\lambda_1\lambda_2}(\vec k) 
    = \sum _{ \lambda_1',\lambda_2'} 
    \int \frac{d^3 \vec k'}{m_r} 
    \  \frac{x'(1-x')}{A(k')} 
\nonumber\\ \times
    U_{\lambda_1\lambda_2;\lambda_1'\lambda_2'}
    (\vec k;\vec k')
    \ \psi_{\lambda_1'\lambda_2'}
    (\vec k')  
\\ +
    \left(E _1(k) + E _2(k)\right)^2
    \psi_{\lambda_1\lambda_2}(\vec k)  
.\nonumber\end{eqnarray} 
Due to the Jacobian, the effective kernel 
\begin{equation}
    \frac{x'(1-x')}{A(k')}\ U(\vec k;\vec k') 
\end{equation} 
ceases to be symmetric in $\vec k$ and $\vec k'$. 
The asymmetry, however, can be
removed by introducing the reduced wave function 
$\varphi_{s_1 s_2}(\vec k )$ defined by 
\begin{eqnarray} 
 \begin{array}{@{}l@{\ }c@{\ }l@{}} 
    \displaystyle 
    \psi_{\lambda_1\lambda_2}(\vec k) 
    &=& 
    \displaystyle 
    \frac{\Phi_{\lambda_1\lambda_2}(\vec k)}{\sqrt{x(1-x)}} ,
\\  
    \displaystyle 
    \Phi_{\lambda_1\lambda_2}(\vec k) 
    &=&     
    \displaystyle 
    \sqrt{A(k)} \sum _{s_1, s_2}
    \Omega _{\lambda_1\lambda_2;s_1 s_2} 
    \ \varphi_{s_1 s_2}(\vec k) .
 \end{array}  
\label{eq:19}\end{eqnarray}   
The equation for $\varphi_{s_1 s_2}$ is then 
\begin{eqnarray}
    &&\left[M^2-\left(E _1(k) + E _2(k)\right)^2\right] 
    \varphi _{s_1 s_2}(\vec k) 
\nonumber\\ &=&
    \sum _{s'_1, s'_2} 
    \!\int\!\! d^3 \vec k' 
    \ \widetilde U _{s_1 s_2;s'_1 s'_2} (\vec k;\vec k')
    \ \varphi_{s'_1 s'_2}(\vec k')  
.\label{eq:18}\end{eqnarray} 
The kernel, with $m_s=m_1+m_2$, is
\begin{eqnarray}
 \begin{array}{@{}l@{}c@{}l@{}} 
    \displaystyle 
    \widetilde U _{s_1 s_2 ;s'_1 s'_2}  
    &=& 
    \displaystyle 
    -
    \frac{4m_s}{3\pi^2}
    \frac{\overline\alpha(Q)}{Q^2}R(Q)
    \frac{\widetilde S _{s_1 s_2;s'_1 s'_2}}
    {\sqrt{A(k) A(k')}} ,
 \end{array}  
\end{eqnarray} 
with the BD spinor factor $\widetilde S $
given in Eq.(\ref{eq:11}).

\section{Doing even better}
The asymmetry can also be
removed by introducing another reduced wave function 
$\overline \varphi_{T S}(\vec k )$,
{\it i.e.} replacing the $\Phi$ in Eq.(\ref{eq:19}) by 
\begin{eqnarray} 
 \begin{array}{c@{}l@{}l@{}} 
    \Phi_{\lambda_1\lambda_2}(\vec k) 
    &=& \displaystyle
    \sqrt{A(k)} \displaystyle \sum _{T,S}
    \big(\Omega C \big) _{\lambda_1\lambda_2;T S} 
    \overline \varphi_{T S}(\vec k) .
 \end{array}  
\end{eqnarray}   
The equation for $\overline \varphi_{T S}$ becomes then
\begin{eqnarray}
    &&\left[M^2-\left(E _1(k) + E _2(k)\right)^2\right] 
    \overline \varphi _{T S}(\vec k) 
\nonumber\\ &=&
    \sum _{T' S'} 
    \!\int\!\! d^3 \vec k' 
    \ \overline U _{T S ;T' S'} (\vec k;\vec k')
    \ \overline \varphi_{T'S'}(\vec k')  
,\end{eqnarray} 
with the kernel ($m_s=m_1+m_2$)
\begin{eqnarray}
 \begin{array} {@{}l@{}l@{}l@{}l@{}l@{}}
    \overline U _{T S ;T' S'}  &=& -
    \displaystyle 
    \frac{4m_s}{3\pi^2}
    \frac{\overline\alpha(Q)}{Q^2}R(Q)
    \frac{\overline S _{T S ;T'S'} }
    {\sqrt{A(k) A(k')}}
 \end{array}  
.\label{eq:24}\end{eqnarray} 
Since $\overline U _{T S ;T' S'} $ is block diagonal, 
as shown in the appendix,
one can divide the set of 4 equations
into an uncoupled equation for the singlets
\begin{eqnarray}
    &&\left[M^2-\left(E _1(k) + E _2(k)\right)^2\right] 
    \overline \varphi _{00}(\vec k) 
\nonumber\\ &=&
    \!\int\!\! d^3 \vec k' 
    \ \overline U _{00} (\vec k;\vec k')
    \ \overline \varphi_{00}(\vec k')  
,\end{eqnarray} 
and a set of 3 coupled equations for the triplets.
They look like Eq.(\ref{eq:24}) except that
the sum over $T=T'=1$ is absent.
The singlet kernel is 
\begin{eqnarray}
 \begin{array} {@{}l@{}l@{}l@{}l@{}l@{}}
    \overline U _{00}  &=& -
    \displaystyle 
    \frac{4m_s}{3\pi^2}
    \frac{\overline\alpha(Q)}{Q^2}R(Q)
    \frac{\overline S _{00} }
    {\sqrt{A(k) A(k')}}
 \end{array}  
,\end{eqnarray} 
with singlet spinor function given in Eq.(\ref{eq:49}).
For equal masses $m_1 = m_2$, it reduces to
\begin{eqnarray} 
 \begin{array} {@{}l@{}l@{}l@{}l@{}l@{}}
 \overline  S _{00} &=& \overline N 
                     [  1 &+& 3\vec k ^{\,2} + 3\vec k^{\prime 2} 
		          +  \vec k ^2 \vec k ^{\prime 2}  
		          -  (\vec k \vec k')(\vec k \vec k') ] ,
 \end{array} 
\label{eq:25}\end{eqnarray}
with the famous hyperfine coefficient 3.
One can thus calculate eigenvalues and eigenfunctions
for the singlets separately from those of the triplets!

It is possible to simplify the triplet equations even further,
by coupling (in the spinor function) the total spin $\vec T$
with the (intrisic) orbital angular momentum $\vec \ell$
to good total momentum $\vec J= \vec \ell + \vec T$,
but this exceeds the limitations of these proceedings, 
and will be given elsewhere.

\section{Conclusion}
We seem to have got the cookie \underline{and} the cake.

\section{Acknowledgements}
AK was supported by the Max--Planck Institut f\"ur Kernphysik, 
Heidelberg, by EU contract number HPCF-CT-2001-00385
with the European Commission High-Level Scientific Conferences for
the workshop TRENTO 2001 as a young researcher, and by Austrian
\textsl{FWF} research grant Nr.~P14794.--
HCP thanks with pleasure Tobias Frederico from Sao Paulo
for verifying independently some of the results.

\begin{appendix}
\section{Spinors and currents}
The Bj\o rken-Drell spinor $u$ is
\begin{equation} 
   u(p,s) = \sqrt{\frac{E+m}{2m}} \left( 
   \begin{array}{@{}r@{}r@{}}\chi_{s} \\ \displaystyle
   \frac{\vec\sigma\cdot\vec p}{E+m}\ \chi_{s} 
   \end{array} \right) 
,\end{equation} 
with the Pauli spinors $\chi_{\uparrow\downarrow}$.
In compact notation  
\begin{eqnarray} \begin{array}{r@{}c@{}cr@{}c@{}cc}
   \vec k_1 &=& \displaystyle\frac{\vec p}{E_1+m_1} , &
   \vec k_2 &=& \displaystyle\frac{\vec p}{E_2+m_2} , \\ \\
   N_1 &=& \sqrt{\displaystyle\frac{E_1+m_1}{2m_1} }, &
   N_2 &=& \sqrt{\displaystyle\frac{E_2+m_2}{2m_2} },
\end{array} \end{eqnarray} 
the spinors become for $\vec p_1=-\vec p_2=\vec p$
\begin{eqnarray} 
\begin{array}{rcc@{}r@{}ccccc} 
   u(p_1,s_1) &=& N_1 & 
   \left(\begin{array}{@{}r@{}}
     \chi_{s_1} \\ \phantom{-}\vec\sigma\cdot\vec k_1\ \chi_{s_1} 
   \end{array} \right) , 
\\ \\
   u(p_2,s_2) &=& N_2 & 
   \left(\begin{array}{@{}r@{}}
     \chi_{s_2} \\ -\vec\sigma\cdot\vec k_2\ \chi_{s_2} 
   \end{array} \right)  .
\end{array} 
\end{eqnarray}
The time-component of the current becomes 
\begin{eqnarray}
 \begin{array} {l@{}l@{}l}
   &\left[ \overline u (p,s)\gamma^0 u(p',s')\right] = \\
   &N N' \langle s\vert 1 + 
   (\vec\sigma\cdot\vec k) 
   (\vec\sigma\cdot\vec k\,') \vert s'\rangle .
 \end{array} 
\end{eqnarray} 
Since 
\(
   (\vec \sigma\cdot \vec k )\ (\vec \sigma\cdot \vec k\,') =
   \vec k \cdot \vec k\,' + i\vec \sigma\cdot (\vec k \wedge \vec k\,')
\)
we abbreviate
\begin{eqnarray}
 \begin{array} {llll}
   K = 1 + \vec k \cdot \vec k\,' ,
   &\hspace{3em}
   &\vec R = i(\vec k \wedge \vec k\,') , 
 \end{array}
\label{eq:26}\end{eqnarray} 
to get in compact notation 
\begin{eqnarray}
 \begin{array} {@{}l@{}l@{}l@{}}
   &\left[ \overline u (p,s)\gamma^0 u(p',s')\right]  
   = N N' 
   \langle s\vert K +
   \vec \sigma \cdot \vec R 
   \vert s'\rangle  
\\
   =& 
   N N' \left(
   \begin{array}{@{}rr@{}} 
     K + \vec e_z\cdot \vec R & \vec e_l\cdot \vec R \\ 
     \vec e_r\cdot \vec R     &- K - \vec e_z\cdot \vec R 
   \end{array}
   \right) ,
 \end{array} 
\label{eq:30}\end{eqnarray}
with $\vec e _l\equiv \vec e _x -i\vec e _y$ 
and $\vec e _r\equiv \vec e _x +i\vec e _y$
and the unit vectors in the three space directions  
$\vec e _x,\ \vec e _y,\ \vec e _z$. 
\\ The space-component of the current becomes 
\begin{eqnarray}
 \begin{array} {l@{}l@{}l}
   \left[ \overline u (p,s)\vec \gamma u(p',s')\right] = \\
   N N'\langle s\vert
   (\vec\sigma\cdot\vec k )\vec \sigma  
   + 
   \vec \sigma (\vec\sigma\cdot\vec k\,') 
   \vert s'\rangle .
 \end{array} 
\end{eqnarray}
With 
\(
   (\vec k \cdot \vec \sigma ) \vec \sigma  =
   \vec k + i(\vec \sigma \wedge \vec k )
\)
and correspondingly 
with
\(
   \vec \sigma \ (\vec k\,' \cdot \vec \sigma ) =
   \vec k\,' - i(\vec \sigma \wedge \vec k\,' )
,\)
we abbreviate 
\begin{eqnarray}
 \begin{array} {llll}
   \vec S = \vec k +\vec k\,' , \qquad
   &\hspace{2em}
   &\vec D =i(\vec k -\vec k\,')
 \end{array} 
,\label{eq:29}\end{eqnarray} 
to get in compact notation 
\begin{eqnarray}
 \begin{array} {@{}l@{}l@{}l@{}}
   &\left[ \overline u (p,s)\vec\gamma\, u(p',s')\right]   
   = N N'\langle s\vert 
   \vec S + \vec \sigma \wedge \vec D\vert s'\rangle  
\\
   =&  
   N N' \left(
   \begin{array}{@{}rr@{}} 
     \vec S + \vec e_z\wedge \vec D & \vec e_l\wedge \vec D \\ 
     \vec e_r\wedge \vec D     &-\vec S - \vec e_z\wedge \vec D 
   \end{array}
   \right) .
 \end{array} 
\label{eq:33}\end{eqnarray}
It helps to evaluate the matrix elements of $\widetilde S$.

\section{The spinor factor $\widetilde S$}
We want to calculate the spinor factor 
\begin{eqnarray} 
 \begin{array} {l@{}l@{}l@{}}
   &&\widetilde S_{s_1 s_2; s_1' s_2'}(\vec p;\vec p\,') 
\\ 
   &=&
   [\overline u (p_1,s_1)\gamma^\mu u(p_1',s_1')] 
   [\overline u (p_2,s_2)\gamma_\mu u(p_2',s_2')]  
\\
   &=&
   [\overline u (p_1,s_1) \gamma^0 u(p_1',s_1')] 
   [\overline u (p_2,s_2) \gamma^0 u(p_2',s_2')] 
\\  
   &-&
   [\overline u (p_1,s_1) \vec\gamma\ \,u(p_1',s_1')] 
   [\overline u (p_2,s_2) \vec\gamma\ \,u(p_2',s_2')] 
 \end{array} 
\label{eq:31}\end{eqnarray}
and arrange it for this purpose as the matrix
\begin{eqnarray} 
  \left[ 
  \begin{array}{@{}c@{\,}c@{\,}c@{\,}c@{}}
    \scriptstyle \widetilde S _{\uparrow\downarrow;\uparrow\downarrow} & 
    \scriptstyle \widetilde S _{\downarrow\uparrow;\uparrow\downarrow} & 
    \scriptstyle \widetilde S _{\uparrow\uparrow;\uparrow\downarrow} & 
    \scriptstyle \widetilde S _{\downarrow\downarrow;\uparrow\downarrow}  
  \\ 
    \scriptstyle \widetilde S _{\uparrow\downarrow;\downarrow\uparrow} & 
    \scriptstyle \widetilde S _{\downarrow\uparrow;\downarrow\uparrow} & 
    \scriptstyle \widetilde S _{\uparrow\uparrow;\downarrow\uparrow} & 
    \scriptstyle \widetilde S _{\downarrow\downarrow;\downarrow\uparrow}   
  \\ 
    \scriptstyle \widetilde S _{\uparrow\downarrow;\uparrow\uparrow} & 
    \scriptstyle \widetilde S _{\downarrow\uparrow;\uparrow\uparrow} & 
    \scriptstyle \widetilde S _{\uparrow\uparrow;\uparrow\uparrow} & 
    \scriptstyle \widetilde S _{\downarrow\downarrow;\uparrow\uparrow} 
  \\ 
    \scriptstyle \widetilde S _{\uparrow\downarrow;\downarrow\downarrow} & 
    \scriptstyle \widetilde S _{\downarrow\uparrow;\downarrow\downarrow} & 
    \scriptstyle \widetilde S _{\uparrow\uparrow;\downarrow\downarrow}& 
    \scriptstyle \widetilde S _{\downarrow\downarrow;\downarrow\downarrow}  
 \end{array} 
 \right] = \left[ 
  \begin{array}{@{}c@{\,}c@{\,}c@{\,}c@{}}
    \scriptstyle \widetilde S _{11} & 
    \scriptstyle \widetilde S _{21} & 
    \scriptstyle \widetilde S _{31} & 
    \scriptstyle \widetilde S _{41}  
  \\ 
    \scriptstyle \widetilde S _{12} & 
    \scriptstyle \widetilde S _{22} & 
    \scriptstyle \widetilde S _{32} & 
    \scriptstyle \widetilde S _{42}  
  \\ 
    \scriptstyle \widetilde S _{13} & 
    \scriptstyle \widetilde S _{23} & 
    \scriptstyle \widetilde S _{33} & 
    \scriptstyle \widetilde S _{43}  
  \\ 
    \scriptstyle \widetilde S _{14} & 
    \scriptstyle \widetilde S _{24} & 
    \scriptstyle \widetilde S _{34} & 
    \scriptstyle \widetilde S _{44}  
 \end{array} 
 \right] .  
\end{eqnarray} 
Its elements become with the above abbreviations   
\begin{eqnarray} 
 \begin{array} {l@{}l@{}l@{}}
   &&\widetilde S_{s_1 s_2; s_1' s_2'}(\vec p;\vec p\,') 
\\
   &=&
   \overline N \langle s_1\vert K  +
   \vec \sigma \cdot \vec R
   \vert s'_1\rangle 
   \langle s_2\vert K  +
   \vec \sigma \cdot \vec R
   \vert s'_2\rangle 
\\ 
   &+&
   \overline N \langle s_1\vert 
   \vec S +  \vec \sigma \wedge \vec D 
   \vert s'_1\rangle
   \langle s_2\vert 
   \vec S +  \vec \sigma \wedge \vec D 
   \vert s'_2\rangle , 
 \end{array} 
\end{eqnarray} 
where $\overline N \equiv N_1 N_1' N_2 N_2'$. 
Prior to explicit calculation, we mention the obvious symmetries
of Eq.(\ref{eq:31}), namely (1) hermiticity, and
(2) symmetry under the particle exchange $1\leftrightarrow 2$.
Since the momenta are back to back,
one can restrict to 
$s_1,s_2 \leftrightarrow s_2,s_1 $,
to spin exchange, which generates the symmetries:
$\widetilde S _{\uparrow\downarrow;\uparrow\downarrow} =
 \widetilde S _{\downarrow\uparrow;\downarrow\uparrow} $, 
$\widetilde S _{\uparrow\downarrow;\downarrow\uparrow} =
 \widetilde S _{\downarrow\uparrow;\uparrow\downarrow} $,
$\widetilde S _{\uparrow\downarrow;\uparrow\uparrow} =
 \widetilde S _{\downarrow\uparrow;\uparrow\uparrow} $, 
and    
$\widetilde S _{\uparrow\downarrow;\downarrow\downarrow} =
 \widetilde S _{\downarrow\uparrow;\downarrow\downarrow} $.
Together with  
$\widetilde S _{\uparrow\uparrow;\uparrow\uparrow} =
 \widetilde S _{\downarrow\downarrow;\downarrow\downarrow} $, 
one has the five relations
\begin{eqnarray} 
 \begin{array} {@{}l@{}l@{}l@{}l@{}l@{}l@{}l@{}l@{}l@{}l@{}l@{}l@{}l@{}l
 @{}l@{}l@{}l@{}l@{}l@{}l@{}l@{}}
 \widetilde S _{11} &=&
 \widetilde S _{22} &;\qquad&
 \widetilde S _{12} &=&
 \widetilde S _{21} &;&
\\
 \widetilde S _{13} &=&
 \widetilde S _{23} &;&
 \widetilde S _{14} &=&
 \widetilde S _{24} &;\quad&
 \widetilde S _{44} &=&
 \widetilde S _{33} &.&
 \end{array} 
\label{eq:34}\end{eqnarray} 
The real non-trivial matrix elements become: 
\begin{eqnarray} 
 \begin{array} {@{}l@{}l@{}l@{}l@{}l@{}l@{}l@{}l@{}l@{}l@{}l@{}l@{}l@{}l
 @{}l@{}l@{}l@{}l@{}l@{}l@{}l@{}}
 \widetilde S _{11} &=& K_1 K_2 
                    &+& \vec S _1\vec S _2 &-& \vec D _1\vec D _2 
                    &-& R _{1z} R _{2z}    &+& D _{1z} D _{2z} 
\\
 \widetilde S _{12} &=&  
                    &+& \vec R _1\vec R _2 &+& \vec D _1\vec D _2 
                    &-& R _{1z} R _{2z}    &+& D _{1z} D _{2z} 
\\
 \widetilde S _{33} &=&K_1 K_2 
                    &+& \vec S _1\vec S _2 &+& \vec D _1\vec D _2 
                    &+& R _{1z} R _{2z}    &-& D _{1z} D _{2z} .
 \end{array}\!\!\!\!\!\! 
\label{eq:35}\end{eqnarray}
The overall factor $\overline N$ is omitted here 
to fit the expressions in one line.
The non-trivial off-diagonal matrix elements
are the complex functions:
\begin{eqnarray} 
 \begin{array}  {@{}l@{}l@{}l@{}}
 \widetilde S _{13} &=& (K_1 + R _{1z})R _{1r} - D _{1z} D _{2r}
                     - (\vec S _1\wedge \vec D _2) _{z}  
\\
 \widetilde S _{14} &=& (K_1 - R _{1z})R _{2l} - D _{1z} D _{2r}
                     - (\vec S _1\wedge \vec D _2) _{l}  
\\
 \widetilde S _{34} &=& R_{1l} R_{2l} - D_{1l}D_{2l}    ,
 \end{array}\!\!\!\!\!\! 
\end{eqnarray} 
where for example $R_{1r}= \vec e_r\cdot \vec R _1$.~--
We like to demonstrate the calculation at hand of two examples.
One begins with inserting the spin projections into Eq.(\ref{eq:31}).
For $S_{12}=\widetilde S_{\uparrow \downarrow; \downarrow \uparrow }$  
one has
\begin{eqnarray} 
 \begin{array} {@{}l@{}l@{}l@{}c@{}l@{}l@{}l@{}l@{}}
   \widetilde S_{12} 
   &=&\overline N &&
   \langle\uparrow  \!\vert K _1  +
   \vec \sigma \cdot \vec R _1
   \vert\!\downarrow\rangle 
   \langle\downarrow\!\vert K _2 +
   \vec \sigma \cdot \vec R _2
   \vert\!\uparrow  \rangle 
\\ 
   &+& \overline N &&
   \langle \uparrow\!\vert 
   \vec S _1+  \vec \sigma \wedge \vec D _1
   \vert  \!\downarrow\rangle
   \langle\downarrow\!\vert 
   \vec S _2+  \vec \sigma \wedge \vec D _2
   \vert  \!\uparrow\rangle   .
 \end{array}\!\!\! 
\end{eqnarray} 
Inserting Eqs. (\ref{eq:30}) and (\ref{eq:33}) gives
\begin{eqnarray} 
 \begin{array} {@{}l@{}l@{}l@{}l@{}l@{}l@{}l@{}l@{}}
   \widetilde S_{12} &=& \overline N 
   &[&
   ( \vec e _r \!\!\cdot \!\! \vec R _1 )
   ( \vec e _l \!\!\cdot \!\! \vec R _2 ) +
   ( \vec e _r \!\!\wedge\!\! \vec D _1 )
   ( \vec e _l \!\!\wedge\!\! \vec D _2 )
   ] 
\\
   &=& \overline N &[& 
   ( \vec e _x \!\!\cdot \!\! \vec R _1 ) 
   ( \vec e _x \!\!\cdot \!\! \vec R _2 ) +
   ( \vec e _y \!\!\cdot \!\! \vec R _1 ) 
   ( \vec e _y \!\!\cdot \!\! \vec R _2 ) 
\\
   & & &+&  
   ( \vec e _x \!\!\wedge\!\! \vec D _1 )
   ( \vec e _x \!\!\wedge\!\! \vec D _2 ) +
   ( \vec e _y \!\!\wedge\!\! \vec D _1 )
   ( \vec e _y \!\!\wedge\!\! \vec D _2 ) ]. 
 \end{array} \!\!\!\!
\end{eqnarray} 
From the vector identity
\begin{eqnarray} 
 \begin{array} {@{}l@{}l@{}l@{}l@{}l@{}l@{}l@{}l@{}l@{}l@{}l@{}l@{}l@{}l@{}l
                @{}l@{}l@{}l@{}l@{}l@{}l@{}l@{}l@{}l@{}}
   (\vec A & \wedge & \vec B ) & \cdot &   
   (\vec C & \wedge & \vec D ) & = & 
   (\vec A & \cdot  & \vec C ) 
   (\vec B & \cdot  & \vec D ) & - & 
   (\vec A & \cdot  & \vec D ) 
   (\vec B & \cdot  & \vec C )
 \end{array} 
\label{eq:36}\end{eqnarray} 
follows most directly Eq.(\ref{eq:35}).~---
Continuing with 
$\widetilde S_{11} = S_{\uparrow \downarrow; \uparrow \downarrow } $,
we skip the first step and get:
\begin{eqnarray} 
 \begin{array} {@{}l@{}l@{}l@{}c@{}l@{}l@{}l@{}l@{}}
   \widetilde S_{11} 
   &=& \overline N &[&
   ( K_1 + \vec e_z \cdot \vec R_1 )
   ( K_2 - \vec e_z \cdot \vec R_2 ) 
\\
   &&&+&
   ( \vec S_1 + \vec e_z \wedge \vec D_1 )
   ( \vec S_2 - \vec e_z \wedge \vec D_2 )
   ] . 
 \end{array}\!\!\! 
\end{eqnarray} 
Applying Eq.(\ref{eq:36}) gives directly Eq.(\ref{eq:35}).

\section{The spinor factor $\overline S$}
Transforming unitarily to $\overline S = C ^\dagger \widetilde S C$,
taking the Clebsch-Gordan coefficients from  Table~5.2 of \cite{Edm64},
{\it i.e.}
$C_{s_1 s_2 ; T S }\equiv \langle t_1s_1t_2s_2 \vert T S\rangle$:
\begin{equation}
C_{s_1 s_2 ; T S } =   
 \left[
 \begin{array} {@{}l@{\,}|cc@{\hspace{0.3ex}}c@{\hspace{0.3ex}}c@{}}
                       & 00  & 10  &  1\!\!+\!\!1 & 1\!\!-\!\!1 %
\\ \hline
   \uparrow\downarrow \rule[1.5ex]{0ex}{1ex} & 
   \frac{1}{\sqrt{2}} & \frac{1}{\sqrt{2}} &0&0 
\\ 
   \downarrow\uparrow  & \frac{-1}{\sqrt{2}} & \frac{1}{\sqrt{2}} &0&0 
\\ 
   \uparrow\uparrow    & 0 & 0 & 1 & 0 
\\ 
   \downarrow\downarrow&0 & 0 & 0 & 1
 \end{array}
 \right] ,
\end{equation}
and carrying out the matrix product,  
gives 
\begin{eqnarray} 
  \overline S  = \left[ 
  \begin{array}{@{}c@{\,}c@{\,}c@{\,}c@{}}
    \widetilde S _{11} - 
    \widetilde S _{12} & 
    0 & 
    0 & 
    0  
  \\ 
    0 & 
    \widetilde S _{22} + 
    \widetilde S _{12} & 
    \sqrt{2}\, 
    \widetilde S _{31} & 
    \sqrt{2}\, 
    \widetilde S _{41}  
  \\ 
    0 &  
    \sqrt{2}\, 
    \widetilde S _{13} & 
    \widetilde S _{33} & 
    \widetilde S _{43}  
  \\ 
    0 &  
    \sqrt{2}\, 
    \widetilde S _{14} & 
    \widetilde S _{34} & 
    \widetilde S _{44}  
 \end{array} 
 \right] .  
\end{eqnarray} 
One needs only the symmetry relations in Eq.(\ref{eq:34})
to derive this result.
Denoting the diagonal elements by $\overline S _{TS}$,
one has thus 
$\overline S _{00}=\widetilde S _{11} - \widetilde S _{12}$
and
$\overline S _{10}=\widetilde S _{22} + \widetilde S _{12}$,
or explicitely from Eq.(\ref{eq:35})
\begin{eqnarray} 
 \begin{array} {@{}l@{}l@{}l@{}l@{}l@{}}
 \overline  S _{00} &=& \overline N [ K_1 K_2 
                     +  \vec S _1\vec S _2 
                     -  \vec R _1 \vec R _2 
		    &-& 2 \vec D _1\vec D _2 ] 
\\
 \overline  S _{10} &=& \overline N [ K_1 K_2 
                     +  \vec S _1\vec S _2 
		     +  \vec R _1 \vec R _2
		    &-&  2 R _{1z} R _{2z}
\\
                & & &+&  2 D _{1z} D _{2z} ].  
 \end{array} 
\end{eqnarray}
By inspection, the singlet matrix element $S _{00} $ 
is a rotational scalar.
To get it explicitly, the abbreviations in  
Eqs.(\ref{eq:26}) and (\ref{eq:29}) are restored: 
\begin{eqnarray} 
 \begin{array} {@{}c@{}l@{}l@{}l@{}}
 \overline  S _{00} &=  \overline N 
                     [  (1+\vec k_1 \vec k'_1) (1+\vec k_2 \vec k'_2) 
		     +  (\vec k_1 \wedge \vec k'_1) 
		        (\vec k_2 \wedge \vec k'_2)
                        \!\!\!\!\!\!\!
\\
                      +&(\vec k_1 + \vec k'_1)  
		        (\vec k_2 + \vec k'_2) 
		     + 2(\vec k_1 - \vec k'_1)  
		        (\vec k_2 - \vec k'_2) 
 \end{array} \!\!\!\!\!\!\!
\nonumber\end{eqnarray}
or after evaluating the wedge products 
\begin{eqnarray} 
 \begin{array} {@{}l@{}l@{}l@{}l@{}l@{}}
 \overline  S _{00} &=& \overline N 
                     [  1 &+& 3\vec k_1 \vec k_2 + 3\vec k'_1 \vec k'_2
		           +  (\vec k_1 \vec k_2 )(\vec k'_1 \vec k'_2 ) 
\\
		     & &  &-&(\vec k_1 \vec k'_2 )^2 
		           + (\vec k_1-\vec k_2 )(\vec k'_1-\vec k'_2 ) ] . 
 \end{array} \!\!\!\!\!
\label{eq:49}\end{eqnarray}
For equal masses, with $\vec k_1 = \vec k_2$, this reduces to 
Eq.(\ref{eq:25}).
On remains with the problem to diagonalize 
$\overline  S $ in the triplet space.
\end{appendix}


\begin{thebibliography}{99}

\bibitem{Pau98} 
      H.C.~Pauli,
      Eur. Phys. Jour. {\bf C  7} (1999) 289-303.
      hep-th/9809005.

\bibitem{Pau01d}
     H.C. Pauli,
     hep-ph/0111040

\bibitem{Pau01e} 
     H.C.~Pauli, these proceedings.
     
\bibitem{BPP98} 
     S.J.~Brodsky, H.C.~Pauli, and S.S.~Pinsky, 
     Phys.~Lett.~C (Physics Reports) {\bf301} (1998) 299-486. 
     hep-th/9707455. 

\bibitem{Pau00d}
     H. C. Pauli,
     \textit{A compendium of light-cone quantization},
     Nucl. Phys. B (Proc.~Supp.) \textbf{90} (2000) 259-272.
     hep-ph/0103106.

\bibitem{PaF01} 
     M. Frewer and H.C. Pauli, these proceedings.

\bibitem{BjD64}  
     J.D. Bj\o rken and S.D.~Drell, \\
     \textit{Relativistic Quantum Mechanics},  \\
     McGraw-Hill, New York, 1964.

\bibitem{Mel74}
     H.J. Melosh, \\
     Phys. Rev. \textbf{D9} (1974) 1095-1112. 

\bibitem{Dzi87}
     Z. Dziembowski, \\
     Phys. Rev. \textbf{D37} (1987) 768-777. 

\bibitem{AhS92}
     D.V. Ahluwalia and M. Sawicki, \\
     Phys. Rev. \textbf{D47} (1993) 5161-5168. 

\bibitem{Edm64}
     A.R. Edmonds, \\
     \textit{Angular momentum in quantum mechanics}, \\
     Princeton University Press, 1962.
     
\end{thebibliography}
\end{document}